\documentstyle[prl,aps,epsfig,multicol]{revtex}
\begin{document}
\draft
\title{Motional sidebands and direct measurement of the cooling rate in the resonance fluorescence of a single trapped ion}
\author{Ch. Raab, J. Eschner, J. Bolle, H. Oberst, F. Schmidt-Kaler, R. Blatt}
\address{Institut f\"ur Experimentalphysik, Universit\"at Innsbruck,\\
Technikerstra{\ss}e 25, A-6020 Innsbruck, Austria}
\date{\today}
\maketitle

\begin{abstract}

Resonance fluorescence of a single trapped ion is spectrally
analyzed using a heterodyne technique. Motional sidebands due to
the oscillation of the ion in the harmonic trap potential are
observed in the fluorescence spectrum. From the width of the
sidebands the cooling rate is obtained and found to be in
agreement with the theoretical prediction.

\end{abstract}


% ------------------------ Text-----------------------------

\begin{multicols}{2}

\narrowtext

Since the first preparation of a single atom in a Paul trap and
observation of its resonance fluorescence \cite{Neuhauser},
investigation of this light has revealed a range of unique
properties. Examples are its nonclassical nature \cite{Diedrich}
and the highly nonlinear response, in the form of sudden intensity
jumps, of a multi-level atom to continuous laser excitation
\cite{qjumps}. The fluorescence is, at the same time, a unique
tool for determining the state of the atom. This is particularly
obvious for a single particle where each photon emission marks the
respective projection of the atomic wave function into the final
state of the corresponding transition. It is also of great
interest to study, through its resonance fluorescence, the motion
of a single laser-excited particle, e.g. for investigating laser
cooling schemes or in connection with proposals for quantum state
manipulation or quantum information processing with trapped
particles \cite{qustmanip}.

The spectrum of fluorescence of a motionless two-level atom exhibits an elastic
part (Rayleigh peak) which is $\delta$-correlated with the exciting light, and an
inelastic contribution, the Mollow triplet \cite{Mollow}. While the latter marks
spontaneous transitions between the dressed states of the combined atom-light
quantum system and appears when the light intensity approaches saturation, the
elastic contribution dominates for light levels below saturation. For a free atom
this spectrum is modified by the recoil shift. For a trapped atom the elastic
peak is unshifted, and sidebands appear at the characteristic frequencies of the
motion in the trap \cite{Javanainen,Lindberg,Parkins}, with sizes that depend on
the amplitude of this motion. The spectral width of these sidebands reflects the
effective decay of the motional states of an atom in the trap, i.e.~the rate of
transitions which change this state. In particular, if laser excitation provides
optical cooling of the trapped atom, the sideband width reflects the equilibrium
of heating and cooling transitions in the steady state. While such sidebands have
been observed in the fluorescence spectrum of ensembles of trapped neutral atoms
\cite{Jessen}, their investigation for a single particle and analysis of their
width has not been done so far \cite{PTB}. The measurement of cooling rates is
highly interesting in experiments with cold atoms or ions, in particular when
many levels or several light fields are involved such that an optimal set of
parameters is hard to find solely from theoretical arguments.

In this paper we report on a measurement of the resonance
fluorescence of a single trapped Barium ion which reveals
sidebands of the elastically scattered light due to the various
components of the ion's motion in the trapping potential. From the
width of the sidebands which correspond to one of its vibrational
modes in the Paul trap quasi-potential we deduce the cooling rate
induced by the exciting laser. This method will enable us to
perform detailed studies of motional effects of laser radiation in
situ, i.e.~during the laser excitation without further analysis
tools.

In the experiment, a single Ba$^+$ ion is trapped in a 1~mm
diameter Paul trap. The ion is generated by impact ionization of a
weak thermal Ba atomic beam with an electron beam inside the trap.
The trap is suspended in UHV and driven with a 500~V$_{\rm pp}$
radio frequency signal at $f_{Paul} \approx 19$~MHz. The ion is
laser-cooled by simultaneous excitation on its S$_{1/2}$
$\leftrightarrow$ P$_{1/2}$ and P$_{1/2}$ $\leftrightarrow$
D$_{3/2}$ resonance lines at 493.4~nm and 649.7~nm, respectively
\cite{BaCooling}. See Fig.~1 for the relevant levels of Ba$^+$.
The 493~nm light is produced by a frequency doubled diode laser
with external grating resonator described in Ref. \cite{ApplPhys}.
This laser is frequency stabilized to a Te$_2$ resonance line
666~MHz away from the Ba$^+$ line, by modulation transfer
spectroscopy \cite{MTS}. The 650~nm light is generated by a diode
laser with an external grating resonator, stabilized to an optical
resonator. Both lasers have linewidths well below 100~kHz. The
laser beams are combined on a dichroic beamsplitter before they
are focused into the trap, and both light fields are linearly
polarized. The laser intensities at the position of the ion are in
the range of 200~mW/cm$^2$ (493~nm) and 100~mW/cm$^2$ (650~nm).
The 650~nm laser is close to resonance, the 493~nm laser is
red-detuned by about the transition linewidth ($\Gamma =
15.1$~MHz) for Doppler cooling. A 2.8~Gauss magnetic field which
is orthogonal to both the laser wave vector and the laser
polarization defines a quantization axis and lifts the degeneracy
of the Zeeman sublevels. The precise parameters are determined by
fitting an 8-level Bloch equation calculation to a scan of the
fluorescence intensity {\it vs.} the detuning of the 650~nm laser
\cite{Blochfit}.

\vspace{-3mm}

\begin{center}
\begin{figure}[tbp]
\epsfig{file=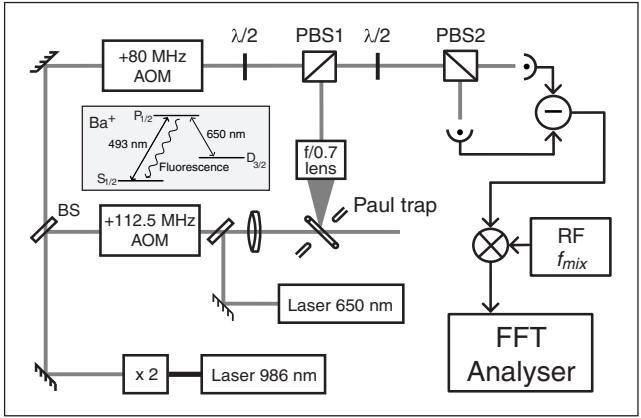,width=0.98\hsize}\vspace{\baselineskip}
\caption{Experimental setup. The inset shows the relevant levels
and transition wavelengths of Ba$^+$. Resonance fluorescence at
493~nm is investigated.~AOM stands for acousto-optic modulator, BS
for beam splitter, PBS for polarizing beam splitter, $\times 2$
for frequency doubling. $f_{mix}$ is varied to observe the various
spectral components of the fluorescence.} \label{Fig. 1}
\end{figure}
\end{center}

The 493 nm resonance fluorescence of the ion is analyzed with a
heterodyne detection setup shown in Fig.~1 \cite{Kimble}. Using a
slightly simpler scheme, with a frequency shift in only one arm of
the heterodyne setup and with a frequency selective photo
detector, H\"offges et al. \cite{Walther} have observed the
elastic part of an ion's resonance fluorescence, but no motional
sidebands. In our setup, the fluorescence at right angle to the
direction of the laser beams, in the direction of the magnetic
field, is collimated with an f/0.7 lens. This light consists of
the two $\sigma$-polarized components corresponding to the
P$_{1/2}(m=\pm 1/2)$ to S$_{1/2}(m=\mp 1/2)$ transitions, and its
elastic part is linearly polarized, because the two
$\sigma$-polarized components are coherently superimposed. For
creating a local oscillator beam, the green laser light is divided
into two beams which are frequency shifted by 112.5~MHz (beam 1)
and 80~MHz (beam 2) using acousto-optical modulators (AOMs). Beam
1 excites the ion while beam 2 is superimposed with the collimated
fluorescence on a polarizing beam splitter PBS1. The beam sizes of
the collimated fluorescence and the local oscillator (beam 2) are
adjusted for optimum overlap, using telescopes. Their orthogonal
polarizations after PBS1 are mixed with a $\lambda$/2 plate and a
second polarizing beam splitter PBS2 to create an interference
signal. The two output signals of PBS2 are detected on two fast
photodiodes (50~MHz bandwidth, 80$\%$ quantum efficiency). By
subtracting their photodiode currents, the interference term $S
\propto E_{lo} E_{fluor}$ between the fluorescence and the local
oscillator is filtered out. Due to the frequency shifts in beams 1
and 2, this interference product (i.e. its elastic part) is
expected at a frequency of 32.5~MHz, the difference frequency of
the two AOM drives, while any motional sideband due to an
oscillatory motion with frequency $f_s$ would appear at 32.5~MHz
$\pm n f_s, n = 1,2,..$. The reason for using two AOMs is that
their difference frequency is not generated elsewhere in the
setup, such that no electrical crosstalk perturbs the final
signal, neither can any residual amplitude modulation in one of
the beams create a 32.5 MHz signal on the photodiodes. The
interference signal $S$ is mixed with an rf reference signal at
variable frequency $f_{mix}$ around 32.5~MHz, low-pass filtered,
and finally analyzed on a Fourier spectrum analyzer with 100 kHz
maximum bandwidth. All rf sources, i.e. the trap drive, the AOM
supplies, and the rf reference, are phase locked to the same
10~MHz master oscillator.

\vspace{-5mm}

\begin{center}
\begin{figure}[tbp]
\epsfig{file=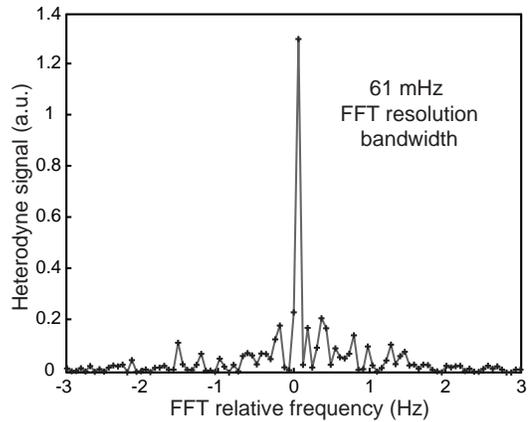,width=0.8\hsize}\vspace{\baselineskip} \caption{FFT
spectrum analyser signal around 50~kHz showing bandwidth-limited heterodyne
detection of the elastic (carrier) peak of the resonance fluorescence.}
\label{Fig. 2}
\end{figure}
\end{center}

With the mixer frequency $f_{mix}$ set to $f_0 = 32.45$~MHz, the
elastically scattered light produces a signal at 50~kHz on the
spectrum analyzer. Such a signal, with the resolution bandwidth
set to 61~mHz, is shown in Fig.~2. Clearly, only one data point is
significantly above the background noise, thus verifying the
$\delta$-correlation between exciting laser and fluorescence. The
signal-to-noise ratio (SNR) is 17 dB (at unit bandwidth). The
maximum SNR which can be achieved depends on the rate $N$ of
fluorescence photons that are detected. It is derived in the
following way: The spectral power of the signal $S$ is $P_S
\propto S^2 /\Delta \nu$, where $\Delta \nu$ is the detection
bandwidth (or signal bandwidth, whichever dominates), while the
noise power is $P_N \propto E_{lo}^2$, such that their ratio is
${\rm SNR}_{max} = E_{fluor}^2 / \Delta \nu = N / \Delta \nu $.
For this to hold, the noise created by $E_{lo}$ with no
fluorescence present \cite{footnote_on_vacuum} has to be the
dominant noise in the photodiode signal $S$. By varying the local
oscillator power without a fluorescence signal we confirmed that
this is the case in our experiment and that the noise is close the
the quantum limit. With a typical scattering rate of $2.5-5 \times
10^4$~photons/s into the solid angle that is collimated, and with
$80\%$ photodiode quantum efficiency, the maximum possible SNR
turns out to be 35-40~dB. The comparatively low value of 17 dB
which we find is due to phase front distortions induced by the
collimating lens and a resulting low degree of mode matching
between the collimated fluorescence and the local oscillator.

Due to the quadrupole radiofrequency field in a Paul trap, the ion
undergoes a driven oscillation at the frequency $f_{Paul} =
18.53$~MHz. This so-called micromotion is in phase with the
driving field, and its amplitude is proportional to the ion's
distance from the trap center. The sidebands to the elastic peak
which this oscillation generates can be observed on the spectrum
analyser, in the same manner as the carrier, by setting $f_{mix}$
to $f_0 \pm n f_{Paul}, n=1,2,..$. In Fig.~3 we show the results
of such measurements. Due to the fact that the phase of the
micromotion is well-defined, and because all rf sources are phase
locked, the width of the micromotion sidebands is limited by the
resolution bandwidth of the spectrum analyser, just like the width
of the elastic peak in Fig.~2.

\vspace{-4mm}

\begin{center}
\begin{figure}[tbp]
\epsfig{file=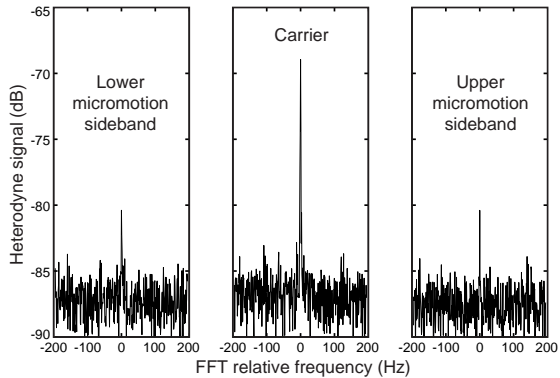,width=0.85\hsize}\vspace{\baselineskip}
\caption{Heterodyne signal showing the elastic peak (center) and
the $n=\pm 1$ micromotion sidebands (right, left). For recording
the sidebands the FFT analyser frequency is shifted by $\pm f_{\rm
Paul} = \pm 18.53$~MHz. The FFT resolution bandwidth was set to
1~Hz. Note the logarithmic ordinate scale.} \label{Fig. 3}
\end{figure}
\end{center}

The sizes of the micromotion sidebands are expected to be
proportional to $|J_n(m)|^2$ where $J$ is a Bessel function,
$n=0,\pm 1, \pm 2,...$ is the number of the sideband, and $m$ is
the modulation index corresponding to the ion's oscillation. With
the micromotion described by $\vec{a} \sin \Omega t$, ($\Omega =
2\pi f_{Paul}$) and with $\vec{k_l}$ and $\vec{k_d}$ describing
the laser and fluorescence wave vectors, respectively, $m$ is
given by $m = \vec{a} \cdot (\vec{k_d} - \vec{k_l})$
\cite{Walther}. The particular value of $m$ found from the
sidebands in Fig.~3 is 0.47. This corresponds to a micromotion
amplitude in the direction of $\vec{k_l}-\vec{k_d}$ of about
26~nm. By minimising the micromotion, i.e. the value of $m$, with
the aid of such a measurement, the ion can be placed in the trap
center where the trapping conditions are optimal \cite{microcomp}.
At optimum conditions, i.e. if the SNR reaches 40~dB, and $\vec{a}
\parallel \vec{k_d} - \vec{k_l}$, this measurement is sensitive to
micromotion amplitudes of about 1~nm.

Apart from the forced micromotion oscillation, the
three-dimensional trapping in the rf-generated pseudo-potential of
a Paul trap results in three independent modes of free vibration
along orthogonal axes at three (in general different) frequencies
$f_{macro}$. In our case, these frequencies are
620.5,~670,~1301~kHz. Investigation of this so-called macromotion
is our main purpose, because these motional degrees of freedom
interact with the laser in a cooling process, and the
corresponding sidebands contain information about the motional
state of the ion, the efficiency of the cooling, and their
dependences on the laser parameters. It is also the macromotion
which is used in experiments on quantum state manipulation and
entanglement, in particular in the framework of quantum
information and quantum computation \cite{state_engineering}.

The macromotion is harder to detect in the heterodyne signal than
the micromotion because it is not correlated with any applied rf
source. In view of the noise considerations above, detection of a
macromotion sideband requires that the rate of photons which
contribute to the sideband heterodyne signal is larger than the
spectral width of that sideband. Due to the inefficient mode
matching described above, this situation is not realized in our
present setup. However, the macromotion corresponds to a damped
harmonic oscillator (laser cooling being the damping mechanism)
which can be driven with an additional external field at some
frequency $f_{drive}$ close to the macromotion frequency
$f_{macro}$. Then, with $f_{mix}$ set to $f_0 \pm n f_{drive}$,
the excited macromotion is detectable as a $\delta$-signal on the
spectrum analyser, and a scan of $f_{drive}$ over one of the
macromotion resonances reveals the response of that oscillator
mode to the drive, in particular the broadening of the resonance
due to laser cooling. The result of such a measurement is shown in
Fig.~4. Here, the upper trace corresponds to the height of the
elastic peak ($f_{mix}=f_{0}$) as in Fig.~2, while the lower trace
corresponds to the height of the sideband at $f_{drive}$
($f_{mix}=f_0+f_{drive}$), both as functions of $f_{drive}$ around
the 620.5~kHz macromotion mode frequency. The driving voltage was
applied to an electrode about 1~mm away from the ion, its power
was -60 dBm. The axis of the excited vibrational mode is at 45$^o$
between the direction of the laser and the observation, as was
found from the image of the ion at much higher drive power.

Fig.~4 shows how the elastic scattering decreases while the
sideband scattering increases around the macromotion resonance.
The two traces were fitted with functions $|A_n
J_n(m(f_{drive}))|^2, n=0 $ for the carrier, $n=1$ for the
sideband, where $m(f_{drive})$ was assumed as
\begin{equation}
m(f_{drive})=\frac{m_{max}}{1+(\frac{f_{drive}-f_{macro}}{\Delta f/2})^2}
\end{equation}
as expected for a damped harmonic oscillator. We find a maximum
modulation index $m_{max}=1.5$ and a width of the macromotion
resonance $\Delta f=750$~Hz. This width corresponds to the
effective linewidth that the ion's oscillator eigenstates in the
trap acquire due to the ongoing laser excitation which makes the
ion change its motional state \cite{Lindberg,Parkins,Walther}.

\vspace{-3mm}

\begin{center}
\begin{figure}[tbp]
\epsfig{file=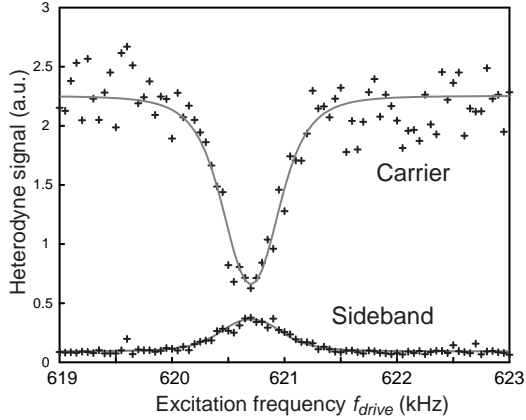,width=0.8\hsize}\vspace{\baselineskip} \caption{Heterodyne
signal showing the sizes of the elastic peak (upper trace) and the sideband of
the weakly excited macromotion (lower trace) as functions of the excitation
frequency $f_{\rm drive}$. For the elastic peak the FFT analyser frequency is
kept fixed while for the sideband it is shifted by $f_{\rm drive}$. The bandwidth
is 1~Hz.} \label{Fig. 4}
\end{figure}
\end{center}

To compare this cooling rate with the one derived from a simple
model, we describe the motion of the ion by a driven and damped
harmonic oscillator. Some limiting conditions are fulfilled in our
experiment: The cooling rate is much smaller than the linewidth of
the transition $\Gamma$, i.e. the velocity changes only negligibly
during one lifetime, and both the recoil frequency $\hbar
k^2/2M=2\pi\times 5.9$~kHz, $M$ being the ion mass, and the
maximum Doppler shift $kv_{max}$ are smaller than the oscillation
frequency (the latter being the Lamb-Dicke condition). Therefore
we can calculate the cooling rate, i.e.~the damping coefficient,
from the radiation pressure that acts on the ion during its
oscillation \cite{coolingbasics}
\begin{equation} F(v)= \hbar k \Gamma P_{\rm P}(\Delta - k v),
\end{equation}
where $P_{\rm P}$ is the probability of finding the ion in the
P$_{1/2}$ state and $\Delta$ is the detuning of the 493~nm laser
from the atomic resonance frequency. With $\Delta$ negative and of
the order of $\Gamma$, and for $kv \ll \Gamma,|\Delta|$ the
velocity-dependent part of $F(v)$ amounts, in first order, to a
friction force $F_f=-\alpha Mv$ that the ion experiences
\cite{coolingbasics}, and
\begin{equation}
\alpha = 2 \frac{\hbar k^2}{2M} \Gamma \frac{dP_{\rm P}}{d\Delta}
\end{equation}
corresponds to the linewidth induced by the laser cooling. All
parameters in Eq.(3) refer to the 493~nm laser. We can neglect the
contribution of the 650~nm laser to the cooling because this laser
is set to yield maximum fluorescence ($\Delta_{650}= 2\pi\times
5$~MHz), such that $dP_{\rm P}/d\Delta_{650} \approx 0$. By
calculating $dP_{\rm P}/d\Delta$ from the same 8-level Bloch
equations that determine our experimental parameters
\cite{Blochfit} $\Delta_{493} = -2\pi\times 19$~MHz,
$I_{493}=189~{\rm mW/cm}^2$, $I_{650}=107~{\rm mW/cm}^2$, we get
$\alpha=2\pi\times 640$~Hz which agrees well with the measurement.

In conclusion, the spectrum of resonance fluorescence from a single harmonically
confined ion in a Paul trap was observed using heterodyne detection. Aside from
the elastic peak, the spectrum exhibits sidebands due to the weakly excited
macromotion of the ion in the trap. Investigation of these sidebands allows for
an analysis of the cooling rate, without affecting the motional state of the ion
by, e.g., probing pulses. The measured cooling rate in our experiment compares
well with a simple model calculation. Moreover, micromotion sidebands are
observed which, by minimizing their amplitude, can be used to actively compensate
for residual micromotion down to the nm-level. These techniques will prove useful
in the context of precision spectroscopy with ion traps and in particular for the
preparation and manipulation of vibrational quantum states of motion as they are
required for quantum information experiments with trapped ions.

We acknowledge support by the Fonds zur F\"orderung der wissenschaftlichen
Forschung (FWF) (project P11467-PHY) and by the EC (TMR network "Quantum
Structures", ERB-FMRX-CT96-0077).

%****************************************************************************%

\end{multicols}


\begin{references}
\bibitem{PTB}
Preliminary results have been reported by B\"uhner and Tamm of
PTB, Braunschweig, Germany .
\end{references}
\end{document}